\documentclass[preprint,showkeys,preprintnumbers,amsmath,amssymb]{revtex4}

\usepackage{graphicx}
\usepackage{dcolumn}
\usepackage{bm}

\def\e{\mathrm{e}}

\begin{document}

\author{R. Rossi Jr.}
\email{romeu.rossi@ufv.br}
\affiliation{Universidade Federal de Vi{\c c}osa - Campus Florestal,
LMG818 Km6, Minas Gerais, Florestal 35690-000, Brazil}

\title{The dynamical quantum Zeno effect in quantum decision theory}

\begin{abstract}
In this paper it is proposed the dynamical quantum Zeno Effect in quantum decision theory. The measurement postulate is not an essential ingredient for the explanation of the quantum Zeno effect, a dynamical account is given in quantum physics. In this account, the entanglement between the system of interest and the apparatus inhibit the quantum transition. The collapse postulate   is not considered. It is show in this paper that the belief-action entanglement model provides a mathematical framework for the dynamical quantum Zeno effect in quantum decision theory. It is also shown that, in this context the dynamical account implies that opinion change process can be inhibited by frequent evaluations of intentions to act. 

\end{abstract}

\maketitle

\section{Introduction}

The Quantum Zeno Effect (QZE) was proposed by B. Misra and E. C. Sudarshan \cite{mis} as a paradox. The theoretical prediction that quantum evolution is frozen when subjected to constant measurements would be in contradiction with experimental observations. In \cite{mis} B. Misra and E. C. Sudarshan hold that the traces of decaying particles in a bubble chambers would be an example of a quantum system that is subjected to constant measurements, however, it continues to evolve. In a later
publication \cite{mis2} they deny the paradoxical character of the QZE. In the case of the of decaying particles in bubble chambers, they conclude that it as not an example of QZE because the observed tracks were not sufficiently frequent to be considered as a continuous measurement. Therefore, the observed tracks could not inhibit the quantum evolution and modify the particle's lifetime.

The first experimental observation of the QZE was reported in 1990 by W. M. Itano et al. \cite{Itano}. They performed an experiment that showed the inhibition of a quantum transition induced by frequent measurements. In the theoretical description of the experiment W. M. Itano et al. hold a controversial interpretation of the QZE, they argue that the projection (or collapse) postulate was an essential ingredient for describing the QZE. Therefore, the QZE would provide experimental evidence for the projection postulate. This interpretation triggered an important debate about the role of the projection postulate in the description of the QZE \cite{Bal,Itano2,Pas}. The approach proposed in the references \cite{Bal,Pas} show that the strong association between the QZE and the projection postulate is not necessary.

In the dynamical approach of the QZE proposed in \cite{Bal,Pas} the intermediate measurements are not represented by protective measurements. This description is replaced by a more complete theory of measurement, in which the interaction between the system of interest and an apparatus is regarded as a quantum process. The measurement process is described by a Hamiltonian quantum evolution, in which the interaction establishes an entanglement between the subsystem of interest and the subsystem associated with the apparatus (auxiliary subsystem). In this dynamical approach of the QZE,  the revelation of the results of each intermediate measurement is irrelevant. The QZE is explained as a consequence of frequent interaction that establishes an entanglement between the subsystem of interest and auxiliary subsystems that represents the apparatus.

The QZE is also addressed in the field of quantum decision theory  \cite{zeno}. In cognitive science the Bayesian probability theory is a commonly used mathematical framework. The decision theory based in this framework is normative, it prescribes logical rules that the rational agent should follow to make decisions. However, several empirical findings have challenge the decision theory that is grounded in Bayesian probability. Two examples are the Prisoner's Dilemma game and the two-stage gambling task \cite{Tve, Tve2, Sha}. Both violate the ``sure thing" principle, which is a fundamental law of Bayesian probability theory \cite{Sav}.

The quantum decision theory \cite{art01,art02,art03,art1,art2,art3,art4} is presented as an alternative framework to be considered in cognitive science. The instrumental use of the quantum probability theory in cognitive science allow the explanation of classical decision making paradoxes \cite{Yuk} and the prediction of results that are in accordance with empirical observations \cite{art01,art02,art03}. It is important to note that in quantum decision theory it is not assumed that the physical brain is a quantum system, the application of quantum theory is merely instrumental.

The authors of reference \cite{zeno} take into account a quantum model to explain an opinion change study. They demonstrate the QZE in this context and provide empirical evidence for it, this corroborates with the quantum decision theory. In the opinion change study \cite{zeno} the participants read a story about a hypothetical murder suspect. The quantum model for this scenario is constructed in a two dimensional Hilbert space in which two orthonormal vectors corresponds to the beliefs that the suspect is either innocent or guilty. Only the participants who believe that the suspect is innocent after reading the initial part of story were selected to the experiment. In the quantum model this preselection corresponds to the preparation of an initial state. The selected participants received pieces of evidences suggesting that the suspect was guilt. In terms of the quantum model, the pieces of evidences yield an unitary evolution of the vector state. This evolution can be characterized as a quantum transition process form the initial ``innocent state" to the final ``guilt state". One group of participants received all the pieces of evidences and were asked to make a final judgment about the suspect's guilt. Other groups of participants received the same pieces of evidences but were also asked to make intermediate judgments in addition to the final judgment. They were asked to express and justify their opinion in each intermediate judgment. The experimental results of \cite{zeno} show that opinion change is affected by intermediate judgments, it slows down this process.

In the quantum model considered in \cite{zeno} the intermediate judgments are represented by the action of POVM operators on the state space, in accordance with the measurement postulate. In the intermediate judgments the participants are asked to express their opinion, therefore, the correspondence with the POVM measurements is well suited. The QZE shown in \cite{zeno} is a result of a sequence of intermediate POVM measurements. However, the QZE can also be obtained dynamically without the application of projective or POVM measurements, as it was shown in \cite{Bal, Pas}.

The purpose of this paper is to study the dynamical description of the QZE in quantum decision theory. A ``gedankenexperiment" is proposed, it is inspired by the opinion change study presented in \cite{zeno}. In this thought experiment the intermediate  judgments expressed by the participants are replaced by intermediate evaluations of actions that are correlated with the participant's beliefs. Neither the evaluations nor its conclusions are to be expressed by the participants. In the quantum model, the evaluations corresponds to the entanglement between two subsystems: $b$ and $a$. The subsystem $b$ is associated with the 
participant's belief and the subsystem $a$ is associated with the participant's intention to act. 
To describe the interaction between $b$ and $a$, we took into account the belief-action entanglement (BAE) model \cite{BAE}. If quantum formalism is suitable for modeling cognitive processes, then the dynamical QZE must also be observed experimentally. This would lead us to the conclusion that a simple evaluation of the intention to act according to one's beliefs is sufficient to slow down the opinion change process, without the need for opinions to be expressed during the process. In the context of opinion change study the dynamical QZE would be an internal process.

In the thought experiment the subsystem $b$ (associated with the participant's belief) is the subsystem of interest. It evolves over time according to the quantum model presented in \cite{zeno}. The  belief-action entanglement (BAE) model \cite{BAE} is used to describe the intermediate interactions. To show how the slowdown of the quantum transition process takes place we consider the effect of a single interaction on the quantum transition process. The dynamical QZE is a result of a sequence of such interactions.

The paper is organized as follows: in section II and III we briefly review the quantum model for the opinion change study presented in \cite{zeno} and the belief-action entanglement (BAE) model given in \cite{BAE}. In section IV we present the thought experiment that allow for the analysis of the dynamical QZE in the quantum decision theory. We show analytically and numerically the effect of a single interaction on the quantum transition process, a sequence of such interactions results in the dynamical QZE.

\section{The Free Evolution}

Let us take into account the quantum model for opinion change studied in \cite{zeno}. The authors consider a two dimensional Hilbert space in which the state vectors are associated with cognitive states of belief. The state vectors are written in the orthonormal basis $\{|0_{b}\rangle, |1_{b}\rangle\}$, where $|0_{b}\rangle$ and $|1_{b}\rangle$ represents the belief that a suspect of the hypothetical murder is either innocent or guilty. The index $b$ identifies the ``belief" subsystem. In the initial stage of the experiment, before presenting the evidences, only the participants who considered the suspect was innocent were selected. This selection corresponds to preparing the initial cognitive state as $|\psi_{0}\rangle =|0_{b}\rangle$. Pieces of evidences suggesting that the suspect was in fact guilty are given to the participants. Therefore, the cognitive state of the participants will evolves to became $|1_{b}\rangle$, which allows us to characterize the process as a quantum state transition.

In a two-level system the vector state evolution can be regarded as a rotation in the Bloch-sphere. In the present case the rotation is governed by the unitary operator 
\begin{equation}
U_{b}(t_{m},t_{n})=\exp(-i\sigma_{x} B(t_{m},t_{n})),
\end{equation}
where $\sigma_{x}$ is the Pauli operator, $t_{j}$ is the time of the presentation of the $j$th piece of evidence, and $B(t_{m},t_{n})$ is the angular displacement of the vector state between the time $t_{m}$ and $t_{n}$, assuming that a judgment  was made at $t_{m}$. The angular displacement function is given by
\begin{equation}
B(t_{m},t_{n})=\alpha\sum^{n}_{j=m+1}a_{j}\exp[-\beta(j-m-1)^{2}],
\end{equation}
where $\alpha$, $\beta$ and $a_{j}$ are parameters that related to the strength of evidences. The details about these parameters are not relevant to the present analysis. The important fact about the this free evolution operator is that the successive applications of $U_{b}(t_{m},t_{n})$ in the vector state $|0_{b}\rangle$ rotates it towards the state $|1_{b}\rangle$. As the form of the time dependence of $B(t_{m},t_{n})$ is not relevant in this work, we use the notation $B(t_{m},t_{n})=B_{m,n}$ and $U_{b}(t_{m},t_{n})=U_{b}(B_{m,n})$.

The judgments of the participants are represented in the quantum model by POVM measurements. 
Quantum theory predicts that a sequence of intermediate POVM measurements can inhibit 
a quantum transition, in \cite{zeno} it is shown that empirical results of opinion change study corroborate with the quantum prediction. It would be an indication that quantum models are useful tools for decision theory.

In \cite{zeno} the authors consider that the intermediate POVM measurements along 
the system's evolution are a necessary ingredient for the QZE. However, as it was 
shown in \cite{Bal, Pas}, the measurement postulate is not necessary for the description
of the QZE. The effect can be achieved by a dynamical analysis, where the intermediate measurements are described by periodically interactions between the system of interest and auxiliary subsystems. The analysis of the dynamical QZE in the context of decision and opinion change is the main purpose of this work.

\section{The Interaction with an Auxiliary Subsystem}

In the section II we presented the free evolution. A state in the subsystem $b$ 
evolves according to the unitary operator $U_{b}(t_{m},t_{n})$. In this case, 
the  participant's belief is the only element relevant in the description. In \cite{BAE} the authors present a model composed by a subsystem associated with beliefs (we call it subsystem $b$) and a subsystem associated with the participant's intention to act (we call it subsystem $a$). The authors of \cite{BAE} studied the violation of sure thing principle of decision theory promoted by the Prisoner’s Dilemma game. To explain the empirical data of this experimental tasks they developed the belief-action entanglement (BAE) model. In BAE model it is assumed that the subject's beliefs interact with the subject's intention to act. This interaction leads to the entanglement between the subsystems $b$ and $a$. In this work we consider that the BAE model allow us to formalize the intermediate evaluations of actions proposed in the thought experiment that is discussed in section IV. This evaluation correlates beliefs and intention to act, in the quantum model they have the same effect as the interactions between the system of interest and the apparatus considered in \cite{Bal,Pas}. These interactions are the essential ingredient for the dynamical QZE.

In the BAE model, the beliefs and the actions are represented by the two dimensional complex subsystems $b$ and $a$. The participant's belief that his opponent will perform the action of cooperation  or the action of defection is characterized by a two level quantum system that is written in the basis $\{|0_{b}\rangle, |1_{b}\rangle\}$, where $|0_{b}\rangle$ and $|1_{b}\rangle$ represents the belief that the opponent will either defect or cooperate, respectively. The index $b$ identifies the ``belief" subsystem. The intention of the action is also characterized by a two level quantum system that is written in the basis $\{|0_{a}\rangle, |1_{a}\rangle\}$, where $|0_{a}\rangle$ and $|1_{a}\rangle$ represents the intention to defect or cooperate with the opponent, respectively. The index $a$ identifies the ``intention to act" subsystem. The global system is composed by the subsystems $b$ and $a$. A vector state in the global system is represented in the basis $\{|0_{b},0_{a}\rangle, |1_{b},0_{a}\rangle, |0_{b},1_{a}\rangle, |1_{b},1_{a}\rangle \}$, where $|k_{b},j_{a}\rangle$ is a short notation of the tensorial product $|k_{b}\rangle \otimes|j_{a}\rangle$ (with $k=0,1$ and $j=0,1$).

The Prisoner’s Dilemma game the players have a period of time to evaluate the pay-offs in order to choose an action. In \cite{BAE} the authors assume that the deliberation process corresponds to an unitary evolution ($U_{ab}$) of a vector state in the global vector space. The Hamiltonian that governs the unitary evolution $U_{ab}$ is given by $H_{ba} = H_{1} + H_{2}$, where:

\begin{equation}
H_{\alpha} = \begin{pmatrix}
\frac{\mu_{0}}{\sqrt{1+\mu_{0}^{2}}} & \frac{\mu_{0}}{\sqrt{1+\mu_{0}^{2}}} & 0 & 0 \\
\frac{1}{\sqrt{1+\mu_{0}^{2}}} & -\frac{\mu_{0}}{\sqrt{1+\mu_{0}^{2}}} & 0 & 0 \\
0 & 0 & \frac{\mu_{1}}{\sqrt{1+\mu_{1}^{2}}} & \frac{1}{\sqrt{1+\mu_{1}^{2}}}\\
0 & 0 & \frac{1}{\sqrt{1+\mu_{1}^{2}}} & -\frac{\mu_{1}}{\sqrt{1+\mu_{1}^{2}}}
\end{pmatrix}
\end{equation}

\begin{equation}
H_{\beta} = \begin{pmatrix}
- \frac{\gamma}{\sqrt{2}}& 0& - \frac{\gamma}{\sqrt{2}} & 0 \\
0 & \frac{\gamma}{\sqrt{2}}& 0 & - \frac{\gamma}{\sqrt{2}} \\
- \frac{\gamma}{\sqrt{2}} & 0 &  \frac{\gamma}{\sqrt{2}}& 0\\
0 & - \frac{\gamma}{\sqrt{2}} & 0& - \frac{\gamma}{\sqrt{2}}
\end{pmatrix}
\end{equation}

The parameters $\mu_{0}$ and $\mu_{1}$ are utility function of the differences in the payoffs: $\mu_{0}=x_{0,0}-x_{0,1}$ and $\mu_{1}=x_{1,0}-x_{1,1}$, where $x_{k,j}$ is the payoff the participant receive if the opponent takes the action $k$ and the participant takes the action $j$. The parameter $\gamma$ is a coupling constant of the interactions promoted by $H_{\beta}$.

The Hamiltonian $H_{\alpha}$ corresponds to an evaluation that is in accord with the maximization of the utility, it is the rational component of $H_{ba}$. The Hamiltonian $H_{\beta}$ allows the possibility for a player's decision not to follow the utility maximization. It implements in the model the tendency of people to change their beliefs to be compatible with their actions. This tendency of human cognition is called ``cognitive dissonance" \cite{Fes}. The combination $H_{\alpha}$ and $H_{\beta}$ results in an evolution more suitable for the analysis of the Prisoner’s Dilemma experiment.

\section{The Dynamical QZE in Quantum Decision Theory}

In this section, it is presented the thought experiment that allow us to study the 
dynamical QZE in quantum decision theory. The thought experiment is similar to the 
of the opinion change study \cite{zeno}. The procedure is the same except for the intermediate judgments. We consider that the participants receive initial information about a murder and a suspect. Those who consider that the suspect is innocent are selected to continue the experiment. After, participants receive pieces of evidence that the suspect is guilty. The participants are separated in two groups. The first group receive all the pieces of evidences 
and in the end are asked to make a judgment. A second group revive the same pieces 
of evidences, but are asked to make intermediate evaluations of actions that are
correlated with the participant's beliefs. For example, they could be asked to 
evaluate is they were willing to report the suspect, or if they would be willing 
to interact socially with the suspect. The possible action must be associated with 
payoffs. Each intermediate evaluation should be different from the previews one, 
in this way they can be represented by different auxiliary subsystems.
Neither the evaluation nor its results can be communicated by the participants. 
In the end of the experiment the participants of the second group also make a final judgment. 
If the quantum model proposed in this paper is suitable to describe the thought experiment, 
the probability that a participant in the second group will judge the suspect as innocent must 
be greater than the same probability for the first group.

The quantum model is constructed in a bipartite Hilbert space 
$\mathbf{H}=\mathbf{H}_{b}\otimes\mathbf{H}_{a}$. The subspace 
$\mathbf{H}_{b}$ is a two dimensional subspace that can be spanned 
by $\{|0_{b}\rangle, |1_{b}\rangle\}$. The subspace $\mathbf{H}_{a}$ 
is also a two dimensional subspace and can be spanned by $\{|0_{a}\rangle, |1_{a}\rangle\}$. In $\mathbf{H}_{b}$ a vector states represents the participant's beliefs about the suspect's guilt or innocence. 
The state $|0_{b}\rangle$ is associated with the belief that the suspect 
is innocent and $|1_{b}\rangle$ is associated with the belief that the 
suspect is guilty. In $\mathbf{H}_{a}$ a vector states represents the participant's 
intention to act. The state $|0_{a}\rangle$ is associated with the intention not to 
act and $|1_{b}\rangle$ is associated with the intention to act. The assumptions about the elements of the vector spaces $\mathbf{H}_{a}$ and $\mathbf{H}_{b}$ are in accordance with the hypothesis of the reference \cite{zeno,BAE}.

In the thought experiment, the initial cognitive state of a 
participant is $|\psi(t_{0})\rangle = |0_{b},0_{a}\rangle$. It indicates the belief that the suspect is innocent and the intention not to report the suspect. To show the dynamical QZE we analyze the consequence of a single intermediate interaction between the subsystems $b$ and $a$. Consider the evolution $U_{I}$ composed by three parts:

\begin{equation}
U_{I}= (U_{b}(B_{1,2})\otimes I_{a})U_{ab}(U_{b}(B_{0,1})\otimes I_{a}),\label{evo}
\end{equation}
where $I_{a}$ is the identity matrix in the subsystem $a$. The index $I$ in $U_{I}$ indicates that the global evolution includes an interaction between the subsystems $b$ and $a$.

The first part is the free evolution $U_{b}(B_{0,1})$ in the subsystem 
$b$ that was considered in \cite{zeno} and discussed in a previous section. 
It rotates towards $|1_{b}\rangle$ the initial state $|0_{b}\rangle$, thus, 
it represents the effect of evidences that the suspect is guilty. It is 
important to emphasize that the subsystem $b$ is supposed to be isolated 
during the evolution $U_{b}(B_{0,1})$. This assumption is necessary to 
represent the state in the belief subsystem $b$ as a pure state, the same 
assumption was made in \cite{zeno}. In the first part of the total evolution 
the subsystem $a$ remains static. If the possibility of act is not mentioned in 
this part of the evolution or before that, the evolution of $a$ and $b$ are expected 
to be independent in the first part of the evolution.

In the second part of the evolution the subsystems $a$ and $b$ interact. We consider that the interaction is governed by the unitary operator $U_{ba}$, given by the belief-action entanglement model \cite{BAE} and discussed in a previous section. In the present case, it corresponds to the participant's evaluation of his intention to act. It is required, for the dynamical explanation of the QZE, that $U_{ba}$ must be able to entangle the subsystems $b$ and $a$. The entanglement between these parts causes decoherence in the subsystem $b$, such decoherence is responsible for the delay in the quantum transition process. This was shown in \cite{Car}.

In the third part the subsystem $b$ evolves freely according to $U_{b}(B_{1,2})$ and the subsystem $a$ does not evolve. The evolution given by factored operator as the one considered in the first part.

The evolution of $|\psi_{b,a}(0)\rangle$ is given by:

\begin{equation}
|\psi_{b,a}(t)\rangle = U_{b}(B_{1,2})U_{ab}U_{b}(B_{0,1})|0_{b},0_{a}\rangle.
\end{equation}

In the first part of the evolution we get:

\begin{eqnarray}
U_{b}(B_{0,1})|\psi_{b,a}(t_{0})\rangle &=& \exp(-i\sigma_{x} B_{0,1})|0_{b},0_{a}\rangle \label{est2}\\
&=&\cos(B_{0,1})|0_{b},0_{a}\rangle - i\sin(B_{0,1}))|1_{b},0_{a}\rangle. \notag
\end{eqnarray}

Notice that $B_{0,1}$ is the rotation angle of the vector in subsystem $b$. 

To make an analytic description of the effect of a single interaction on the free evolution we assume $\gamma=0$. The Hamiltonian that governs the second part of the evolution becomes $H_{\beta\alpha} = H_{\alpha}$. This is an unrealistic assumption, as explained in \cite{BAE}, $\gamma$ is associated with cognitive dissonance tendencies that cannot be suppressed. We assume $\gamma=0$ just to simplify the calculation, this simplification is not an essential ingredient for the dynamical QZE. The numerical analysis with $\gamma \neq 0$ will be presented next.

Under the assumption $\gamma=0$, we have $U_{ba}=\exp(-iH_{\beta\alpha} t^{'}) =\exp(-iH_{\alpha} t^{'})$ where $t^{'}$ is the time of interaction, that corresponds to the time associate with the participant's evaluation of his intention to act. Therefore, the second evolution is given by:

\begin{eqnarray}
U_{ba}(t^{'})|\psi(t_{1})\rangle &=& \exp(-iH_{\alpha} t^{'})\left[\cos(B_{0,1})|0_{b},0_{a}\rangle - i\sin(B_{0,1})|1_{b},0_{a}\rangle\right] \label{est3}\\
&=&\cos(B_{0,1})|0_{b}\rangle |\nu_{a}\rangle - i\sin(B_{0,1})|1_{b}\rangle |\eta_{a}\rangle, \notag
\end{eqnarray}
where

\begin{eqnarray}
|\nu_{a}\rangle &=& \left(\e^{-it^{'}}\cos^{2}\frac{\theta_{0}}{2}+\e^{it^{'}}\sin^{2}\frac{\theta_{0}}{2}\right)|0_{a}\rangle+\left(\e^{-it^{'}}-\e^{it^{'}}\right)\cos\frac{\theta_{0}}{2}\sin\frac{\theta_{0}}{2}|1_{a}\rangle\\
|\eta_{a}\rangle &=& \left(\e^{-it^{'}}\cos^{2}\frac{\theta_{1}}{2}+\e^{it^{'}}\sin^{2}\frac{\theta_{1}}{2}\right)|0_{a}\rangle+\left(\e^{-it^{'}}-\e^{it^{'}}\right)\cos\frac{\theta_{1}}{2}\sin\frac{\theta_{1}}{2}|1_{a}.\rangle\notag
\end{eqnarray}
with $\theta_{0}=\arctan\left(\frac{1}{\mu_{0}}\right)$ and $\theta_{1}=\arctan\left(\frac{1}{\mu_{1}}\right)$. In the present case, the utility functions ($\mu_{0}$ and $\mu_{1}$) can be considered as pay-offs for the possible actions. If the utility functions are $\mu_{0}=\mu_{1}$ we have $|\nu_{a}\rangle=|\eta_{a}\rangle$. In this case the interaction $U_{ba}(t^{'})$ does not entangle the subsystems $b$ and $a$, consequently, it would not slow down the quantum transition process. However, if $\mu_{0}\neq\mu_{1}$ we have $|\nu_{a}\rangle \neq |\eta_{a}\rangle$. The vector state in equation (\ref{est3}) can not be factorized, therefore, it is entangled. It this case, the interaction will cause a slowdown in the quantum transition process in the subsystem $b$. We consider that $\mu_{0}\neq\mu_{1}$.

The third part of the total evolution gives:

\begin{eqnarray}
U_{b}(t_{1},t)|\psi^{'}(t_{1})\rangle &=& \exp(-i\sigma_{x} B_{1,2})\left[\cos(B_{1,2})|0_{b}\rangle |\nu_{a}\rangle - i\sin(B_{1,2})|1_{b}\rangle |\eta_{a}\rangle\right]\\
&=&\cos(B_{0,1}))\left[ \cos(B_{1,2})|0_{b}\rangle-i\sin(B_{1,2})|1_{b}\rangle\right]|\nu_{a}\rangle \notag\\
&&- i\sin(B_{0,1}))\left[ \cos(B_{1,2})|1_{b}\rangle-i\sin(B_{1,2})|0_{b}\rangle\right]|\eta_{a}\rangle, \notag
\end{eqnarray}

The probability that initial state $|0_{b}\rangle$ in the subsystem $b$ is measured after the evolution $U_{I}$ is given by:

\begin{eqnarray}
P^{I}_{0b}(B_{0,1}+B_{1,2})&=&\langle \psi(t_{0})|U^{\dagger}_{I}(|0_{0}\rangle\langle 0_{b}|\otimes I_{a})U_{I}|\psi(t_{0})\rangle\\
&=&\cos^{2}(B_{0,1})\cos^{2}(B_{1,2})+\sin^{2}(B_{0,1})\sin^{2}(B_{1,2})\notag\\
&&-2 \cos(B_{0,1})\cos(B_{1,2})\sin(B_{0,1})\sin(B_{1,2})\operatorname{Re}(\langle \nu_{a}|\eta_{a} \rangle).
\end{eqnarray}
 
It corresponds to the probability that the cognitive state is still consistent with the belief that the suspect is innocent after the evolution $U_{I}$. To make evident the effect induced by entanglement in retaining the initial state in the subsystem $b$, we compare $P^{I}_{0b}$ with the probability $P_{0b}$, that is given by:  

\begin{eqnarray}
P_{0b}(B_{0,1}+B_{1,2})&=&\langle \psi(t_{0})|U^{\dagger}(|0_{0}\rangle\langle 0_{b}|\otimes I_{a})U|\psi(t_{0})\rangle\\
&=&\cos^{2}(B_{0,1})\cos^{2}(B_{1,2})+\sin^{2}(B_{0,1})\sin^{2}(B_{1,2})\notag\\
&&-2 \cos(B_{0,1})\cos(B_{1,2})\sin(B_{0,1}))\sin(B_{1,2}),\notag
\end{eqnarray}
where $U= (U_{b}(B_{1,2})\otimes I_{a})(U_{b}(B_{0,1})\otimes I_{a})$ is a free evolution of the subsystem $b$ without intermediate interaction with the subsystem $a$. Therefore, $P_{0b}(B_{0,1}+B_{1,2})$ is the probability that the cognitive state is still consistent with the belief that the suspect is innocent after the evolution $U$.

The vectors $|\nu_{a}\rangle$ and $|\eta_{a}\rangle$ are unitary, then their inner product results in  $\operatorname{Re}(\langle \nu_{a}|\eta_{a} \rangle)\leq 1$. We conclude that $P_{0b}(B_{0,1}+B_{1,2})\leq P^{I}_{0b}(B_{0,1}+B_{1,2})$, the interaction with the subsystem $a$ slow down the quantum transition process in the subsystem $b$. This effect is shown in FIG.1, where it is clear that the probabilities $P_{0b}(B)$ and $P^{I}_{0b}(B)$ evolve differently after the interaction.

Notice that the slowdown of the opinion change is the result of a mere evaluation of the intention to act. After receiving some pieces of evidence that the suspect is guilty (in FIG.1 $B_{0,1} = 0.2$), the participant is asked about a possibility to act. This action is related to the participant's belief about the murder. The participant only evaluates his intention to act, but does not express his belief or intention to act.

\begin{figure}[h]
\centering
\hspace*{-0.7cm}
  \includegraphics[scale=0.6]{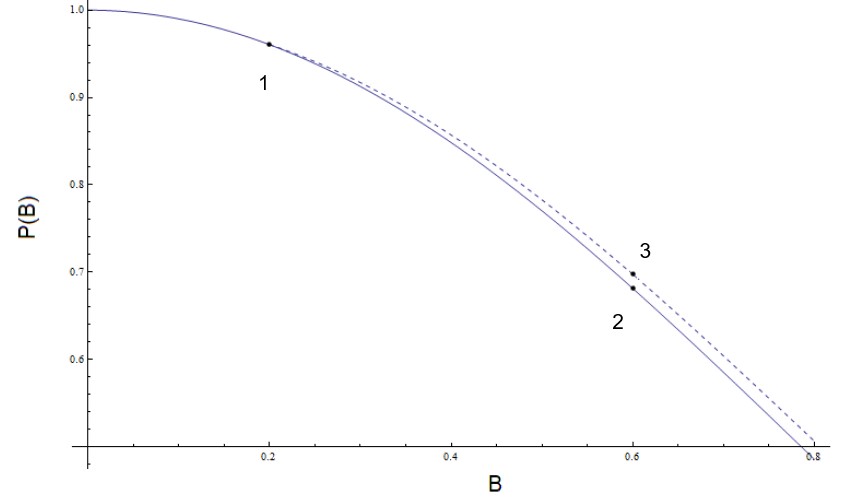}\\
  \caption{$P_{0b}(B) \times B$ (solid line) and $P^{I}_{0b}(B) \times B$, where $B$ is a continuous variable that represents the possible values of $B_{m,n}$. The point $1$ corresponds to $B = B_{0,1} = 0.2$, the points $2$ ($3$) corresponds to $B =B_{0,1}+ B_{1,2} = 0.6$ in $P_{0b}(B)$ ($P^{I}_{0b}(B)$). We consider $\gamma=0$, $\mu_{0}=0.3$, $\mu_{1}=0.6$ and $t^{'}=\pi/2$ }
\end{figure}

In FIG.1 it is shown that the derivative of the probability function $P^{I}_{0b}$ changes after the measurement. The derivative measures the transition rate. Immediately after the measurement it is given by:   
       
\begin{equation}
V_{I}=\lim_{B_{1,2}\rightarrow 0} \frac{d}{dB_{1,2}}P^{I}_{0b}(B_{0,1}+B_{1,2})=-2\cos(B_{0,1})\sin(B_{0,1})\operatorname{Re}(\langle \nu_{a}|\eta_{a} \rangle).
\end{equation}

When no interaction is performed the transition rate is:

\begin{equation}
V=\lim_{B_{1,2}\rightarrow 0} \frac{d}{dB_{1,2}}P_{0b}(B_{0,1}+B_{1,2})=-2\cos(B_{0,1})\sin(B_{0,1}).
\end{equation}

The transition rate is reduced if an interaction takes place: $|V^{'}|\leq|V|$. If the vector $|\nu_{a}\rangle$ is orthogonal to $|\eta_{a} \rangle$ the transition rate $V_{I}$ is null immediately after the interaction. The orthogonality also implies that the state given in equation (\ref{est2}) is a maximally entangled state. In the context of quantum measurement theory, the maximally entangled state characterizes the interaction as a perfect measurement on the system of interest, performed by the auxiliary subsystem. When a sequence of perfect measurements are performed a quantum transition process can be frozen (this is the dynamical QZE), despite the fact that no projections of quantum state are considered \cite{Bal, Pas}.

Notice that the transition rates $V_{I}$ and $V$  depend on the interference terms of the probabilities $P^{I}_{0b}(B_{0,1}+B_{1,2})$ and $P_{0b}(B_{0,1}+B_{1,2})$ respectively. The interaction induces a decoherence effect that reduces the interference terms, consequently, it reduces the transition rate.

If $0<\operatorname{Re}(\langle \nu_{a}|\eta_{a} \rangle)<1$  the interaction results in a partial entangled state. A sequence of these interactions results in a partial QZE, as it was shown in \cite{Per}.

In the case of $-1<\operatorname{Re}(\langle \nu_{a}|\eta_{a} \rangle)<0$ the function $P^{I}_{0b}$ becomes increasing. It means that a subsequent free evolution $U_{b}(B_{1,2})$ will bring the state closer to the initial state. This effect was pointed out in reference \cite{Ros}. This condition will be explored in the context of quantum decision theory in a future work.

\subsection{Dynamics with $\gamma\neq 0$}

In this section we consider model with $\gamma\neq 0$ that corresponds to a more realistic approach. The Hamiltonian that governs the interaction between the subsystems $b$ and $a$ is $H_{\beta\alpha}=H_{\alpha}+H_{\beta}$. It accounts for the utility maximization principle and the cognitive dissonance, as it was discussed in \cite{BAE}.

In FIG.2 it is shown the effect of a single interaction, governed by $H_{\beta\alpha}=H_{\alpha}+H_{\beta}$, in the quantum transition process. The probability $P_{0b}(B)$ decreases continuously while $P^{I}_{0b}(B)$ has a discontinuity. This discontinuity in the graphic is caused by the interaction, which is a dynamical process given by:
\begin{equation}
U_{ba}(t^{'})|\psi(t_{1})\rangle = \exp(-iH_{\beta\alpha} t^{'})\left[\cos(B_{0,1})|0_{b},0_{a}\rangle - i\sin(B_{0,1})|1_{b},0_{a}\rangle\right].     
\end{equation}

The graphic in FIG.2 does not show the evolution of the function $P^{I}_{0b}(B)$ during the interaction. It only shows the result of this dynamical process, which is the discontinuous decrease of $P^{I}_{0b}(B)$. During the interaction process, the probability of measuring the state $|0_{b}\rangle$ in the subsystem $b$ decreases. This discontinuity is not present in the graphic of FIG.1 because during the interaction governed by the Hamiltonian $H_{\alpha}$ the probability of measuring the state $|0_{b}\rangle$ in the subsystem $b$ does not change.

\begin{figure}[h]
\centering
\hspace*{-0.7cm}
  \includegraphics[scale=0.65]{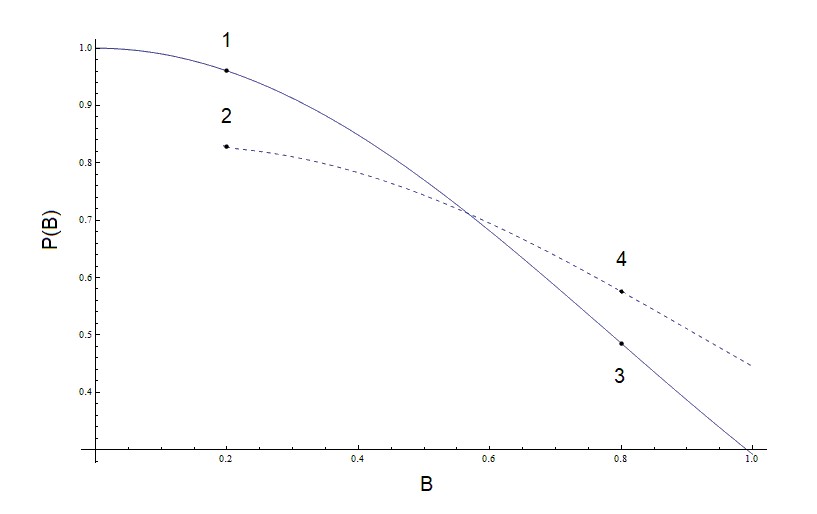}\\
  \caption{$P_{0b}(B) \times B$ (solid line) and $P^{I}_{0b}(B) \times B$, where $B$ is a continuous variable that represents the possible values of $B_{m,n}$. The point $1$($2$) corresponds to $B = B_{0,1} = 0.2$ in $P_{0b}(B)$ ($P^{I}_{0b}(B)$), the points $3$($4$) corresponds to $B =B_{0,1}+ B_{1,2} = 0.8$ in $P_{0b}(B)$ ($P^{I}_{0b}(B)$). We consider $\gamma=2.09$, $\mu_{0}=1.4$, $\mu_{1}=1.4$ and $t^{'}=\pi/2$ }
\end{figure}

We set the parameters $\gamma=2.09$ and $t^{'}=\pi/2$ as it was done in \cite{BAE} to model the results of the Prisoner’s
Dilemma game. We set $\mu_{0}=\mu_{1}=1.4$  to minimize the reduction of the probability $P^{I}_{0b}$ during the interaction. The purpose of this section is to study the quantum transition process governed by the $U_{b}$, and to compare the process with and without intermediate interactions. The increase of the transition probability during the interaction is not the focus of the present work, however, it is an inevitable result of the interaction describe by the Hamiltonian $H_{\beta\alpha}=H_{\alpha}+H_{\beta}$. The setting of the utility function was used to reduce this inevitable effect because these functions have an objective character, therefore, they can be controlled.

To summarize, in this paper it is proposed an analysis of the dynamical QZE in the quantum decision theory. In quantum physics, the dynamical approach of QZE shows that the slowdown of a quantum transition takes place even if no intermediate measurement results are revealed. The dynamical approach of QZE in quantum decision theory implies that an opinion change process is slowed down by frequent evaluations of intentions to act. The actions must be correlated with beliefs, but the mere evaluation is enough to delay the opinion change process. In this sense, the dynamical QZE in the quantum decision theory is associated with an internal process, in which the communication of the subject's beliefs or intentions are not necessary.


\begin{thebibliography}{24}



\bibitem{mis}  B. Misra, E.C. G. Sudarshan, J. Math. Phys. 18 (1977) 756-763.


\bibitem{mis2} C. B. Chiu, E. C. Sudarshan, and B. Misra , Phys. Rev. D 16 (1977) 520-529.


\bibitem{Itano} W. M. Itano, D. J. Heinzen, J. J. Bollinger, D. J. Wineland, Phys. Rev. A 41, 2295 (1990).


\bibitem{Bal} L. E. Ballentine Phys. Rev. A 43, 5165 (1991).


\bibitem{Itano2}  W. M. Itano, D. J. Heinzen, J. J. Bollinger, D. J. Wineland, Phys. Rev. A 43, 5168 (1991).


\bibitem{Pas}  S. Pascazio, M. Namiki Phys. Rev. A 50, 4582 (1994). 


\bibitem{zeno} Yearsley, James M., and Emmanuel M. Pothos. "Zeno's paradox in decision-making." Proceedings of the Royal Society B: Biological Sciences 283.1828 (2016): 20160291.



\bibitem{Tve}  Tversky, A. and Kahneman, D, Judgment under uncertainty: heuristics and biases. Science 185,
1124–1131 (1983).



\bibitem{Tve2} Tversky, A. and Shafir, E,  The disjunction effect in choice under uncertainty. Psychol. Sci. 3, 305–309 (1992).  



\bibitem{Sha} Shafir, E. and Tversky, A,  Thinking through uncertainty:
nonconsequential reasoning and choice. Cogn. Psychol. 24,
449–474 (1992). 




\bibitem{Sav} Savage, L. J, The foundations of statistics. New York, NY: Wiley (1954).




\bibitem{art01} Busemeyer, Jerome R., Peter D. Kvam, and Timothy J. Pleskac. "Markov versus quantum dynamic models of belief change during evidence monitoring." Scientific reports 9.1 (2019): 1-10.



\bibitem{art02} Tsarev, Dmitriy, et al. "Phase transitions, collective emotions and decision-making problem in heterogeneous social systems." Scientific Reports 9.1 (2019): 1-13.


\bibitem{art03} Joong-Sung, Lee, et al. "Experimental Demonstration on Quantum Sensitivity to Available Information in Decision Making." Scientific Reports (Nature Publisher Group) 9.1 (2019).


\bibitem{art1} Busemeyer JR, Bruza P. 2011 Quantum models of cognition and decision making. Cambridge, UK:
Cambridge University Press.



\bibitem{art2} Haven E, Khrennikov A. 2013 Quantum social science. Cambridge, UK: Cambridge University Press.



\bibitem{art3} Pothos EM, Busemeyer JR. 2013 Can quantum probability provide a new direction for cognitive
modeling? Behav. Brain Sci. 36, 255–327. 



\bibitem{art4} Wang Z, Solloway T, Shiffrin RM, Busemeyer JR. 2014 Context effects produced by question orders
reveal quantum nature of human judgments. Proc. Natl Acad. Sci. USA 111, 9431–9436. 



\bibitem{Yuk} Yukalov, Vyacheslav I., and Didier Sornette. "Quantum decision theory as quantum theory of measurement." Physics Letters A 372, 6867-6871 (2008).





\bibitem{BAE} Pothos, Emmanuel M., and Jerome R. Busemeyer. "A quantum probability explanation for violations of ‘rational’decision theory." Proceedings of the Royal Society B: Biological Sciences 276.1665 (2009): 2171-2178.
 
 
\bibitem{Fes} Festinger, L. 1957 A theory of cognitive dissonance. Stanford, CA: Stanford University Press. 
 

\bibitem{Car} A. F. R. de Toledo Piza, and M. C. Nemes. "Manipulating decay rates by entanglement and the Zeno effect." Physics Letters A 290.1-2 (2001): 6-10.


\bibitem{Per} A. Peres and A. Ron, Phys. Rev. A 42, 5720 (1990).



\bibitem{Ros} R. Rossi Jr, A.R.B. de Magalhães, M.C. Nemes Phys. Rev. A 78, 042111 (2008)



\end{thebibliography}
\end{document}